\documentclass[notitlepageprb,showpacs,superscriptaddress,preprint,titlepage,aps,pra]{revtex4-2}
\usepackage{amssymb,amsmath,hyperref}
\usepackage{lineno,hyperref,bm,dcolumn,graphicx}

\begin{document}

\title{Correlation in formation of \texorpdfstring{$^{8}$Be}{8Be} nuclei and \texorpdfstring{$\alpha$}{alpha}-particles in fragmentation of relativistic nuclei}

\author{A.A.~Zaitsev}
\affiliation{Joint Institute for Nuclear Research, Dubna, Russia}
\affiliation{Lebedev Physical Institute, Russian Academy of Sciences, Moscow, Russia}

\author{D.A.~Artemenkov}
\affiliation{Joint Institute for Nuclear Research, Dubna, Russia}

\author{V.V.~Glagolev}
\affiliation{Joint Institute for Nuclear Research, Dubna, Russia}

\author{M.M.~Chernyavsky}
\affiliation{Lebedev Physical Institute, Russian Academy of Sciences, Moscow, Russia}

\author{N.G.~Peresadko}
\affiliation{Lebedev Physical Institute, Russian Academy of Sciences, Moscow, Russia}

\author{V.V.~Rusakova}
\affiliation{Joint Institute for Nuclear Research, Dubna, Russia}

\author{P.I.~Zarubin}
\email[email:]{ zarubin@lhe.jinr.ru}
\affiliation {Joint Institute for Nuclear Research, Dubna, Russia}
\affiliation{Lebedev Physical Institute, Russian Academy of Sciences, Moscow, Russia}

\date{\today}

\begin{abstract}
In the events of peripheral dissociation of relativistic nuclei in the nuclear track emulsion, it is possible to study the emerging ensembles of He and H nuclei, including those from decays of unstable $^{8}$Be and $^{9}$B nuclei, as well as the Hoyle state. These extremely short-lived states are identified by invariant masses calculated from the angles in 2$\alpha$-pairs, 2$\alpha p$- and 3$\alpha$-triplets in the approximation of conservation of momentum per nucleon of the primary nucleus. In the same approach, it is possible to search for more complex states. This paper explores the correlation between the formation of $^{8}$Be nuclei and the multiplicity of accompanying $\alpha$-particles in the dissociation of relativistic $^{16}$O, $^{22}$Ne, $^{28}$Si, and $^{197}$Au nuclei. On the above basis, estimates of this correlation are presented for the unstable $^{9}$B nucleus and the Hoyle state. The enhancement in the $^{8}$Be contribution to dissociation with the $\alpha$-particle multiplicity has been found. Decays of $^{9}$B nuclei and Hoyle states follow the same trend.
\end{abstract}

\pacs{21.60.Gx, 25.75.-q, 29.40.Rg} 

\maketitle 

\section{Introduction}
\label{intro}
The correlated pairs of $\alpha$-particles in decays of $^{8}$Be nuclei can make a noticeable contribution to the final states of nuclear fusion and breakup reactions. These decays are identified by extremely low relative energy of $\alpha$-particles. The lifetime of $^{8}$Be inversely proportional to the width equal to 5.6 eV \cite{Ajzenberg:1988} exceeds the duration of nuclear reactions by several orders of magnitude, that unites the unstable $^{8}$Be nucleus with other fragments. Despite the large distance between the constituent $\alpha$-particles (comparable to the diameter of the Fe nucleus), the unstable $^{8}$Be nucleus is considered as the basis in light nuclei. Upon excitation and fragmentation, their cluster structure is clearly manifested, including the clustered $^{8}$Be nucleus in the ground and first excited states. The exotic dimensions of $^{8}$Be make it a non-trivial probe of the generative reaction dynamics.

A complete study of nuclear clustering in multiple final states relies on a wide range of instruments included in compact spectrometers operating with low-energy nuclear beams (review \cite{Beck:2012}). It also implies the reconstruction of $^{8}$Be decays. The $^{8}$Be nuclei can be produced both in collisions of nuclei and decays of nuclei upon their excitation above the corresponding thresholds. In the latter case, the most significant are the cascade decays of the unstable $^{9}$B nucleus and the Hoyle state (review \cite{Freer:2014}), where $^{8}$Be decays must be present \cite{Smith:2020}. Like $^{8}$Be, each of these states at unusually large sizes has extremely low decay energy, and their lifetime \cite{Ajzenberg:1988} is several orders of magnitude larger than the scale of nuclear interactions. The similarity of these three objects, radically different from the overwhelming majority of nuclei, allows them to be attributed to a special class of unstable states with a nuclear-molecular structure. The ratios of their yields make it possible to characterize the dynamics of reactions in general. The exotic structure HS and $^{9}$B further enhances diagnostic capabilities. However, their statistical security will always be deliberately lower than $^{8}$Be. Therefore, the $^{8}$Be nucleus is the most suitable starting point to study the mechanisms of $\alpha$-particle ensemble formation by the method of unstable states.

It must not be excluded that there are more nuclear-molecular states hidden among nuclear excitations which decay into HS or $^{9}$B and $^{8}$Be. A close candidate is the long-lived excitation of the $^{13}$N isotope at 15.1 MeV (5.6 MeV above the $^{9}$B$\alpha$ or HS$p$ threshold) \cite{Ajzenberg:1988}. Besides, light isotopes usually have long-lived excitations near the $\alpha$-particle separation thresholds, which can be interpreted as paired nuclear-molecular structures where $^{8}$Be is replaced by a stable nucleus, for example, pairs $^{6}$Li$\alpha$, $^{7}$Be($^{7}$Li)$\alpha$, $^{12}$C$\alpha$ and others \cite{Ajzenberg:1988}. The considered energy scale makes it possible to apply the findings in the study of unstable states in nuclear astrophysics.
 
The current interest in $^{8}$Be and HS is due to the success of their description as 2- and 3-body states of the Bose-Einstein $\alpha$-condensate, which appear at the reduced density of the nucleon medium \cite{Tohsaki:2001,Yamada:2004,Schuck:2017}. Following them, the 0$^+_6$ excited state of the $^{16}$O nucleus at 660 keV above the 4$\alpha$ threshold is considered as the 4$\alpha$ analog of HS. Condensates up to 10$\alpha$-particle are to appear. Experimental approaches to search for condensate are offered, including dissociation of relativistic nuclei in the above approach discussed below (review \cite{Oertzen:2010}).

In the context of this study, the observations already made on the search for $\alpha$-condensate are important. The experiment aimed at complete registering $\alpha$-particle fragments of a projectile in the reaction $^{40}$Ca(25 MeV/nucleon) + $^{12}$C has indicated the increase in the $^{8}$Be contribution to $\alpha$-multiplicity of 6. This fact contradicts the model which predicts the decrease (Table 2 in \cite{Borderie:2016}). The focus of experimental efforts was concentrated on the search for $^{16}$O(0$^+_6$) decays in nuclear reactions $^{20}$Ne(12 MeV/nucleon) + $^{4}$He \cite{Barbui:2018} and $^{16}$O (160, 280, 400 MeV) + $^{12}$C \cite{Bishop:2019}. Although the status of observation, and even more so the determination of spin and parity, is still uncertain \cite{Kokalova:2020}, an important conclusion was made that HS(3$\alpha$) is formed during fragmentation not only of $^{12}$C \cite{Barbui:2018,Bishop:2019}. This fact points to the universality of the HS, as well as of the $^{8}$Be nucleus.

Being internally extremely low-energy, the very phenomenon of the condensate state formation should be the relativistic invariant one, i.e., must be presented with increasing energy of the fragmenting nucleus. Moving into the region of the limiting fragmentation of several GeV/~nucleon projectiles allows one to separate the kinematic regions of fragmentation of the incident nuclei and the target. Besides the above provides an additional projection onto unstable states and their surroundings. Due to the collimation of fragments and absence of detection thresholds, there are methodological advantages, which, however, are not easy to use. It is required both - to change the form of representation of $\alpha$-particle correlations to the relativistic invariant one, and use an adequate technique.

The analysis of fragmentation of relativistic nuclei in the nuclear track emulsion (NTE) enables one to study internally non-relativistic ensembles of H and He nuclei produced in decays of unstable $^{8}$Be and $^{9}$B nuclei up to the most complex ones \cite{ZarubinLN:2013,ArtemPPN:2017,ArtemPAN:2017}. NTE layers from 200 to 500 $\mu$m thick, longitudinally exposed to the nuclei under study, enable one, with full completeness and resolution of 0.5 $\mu$m, to determine the angles between the directions of emission of relativistic fragments in the cone $sin(\theta_{fr})$ = $P_{fr}$/$P_{0}$. Here $P_{fr}$ = 0.2 GeV/$c$ is the characteristic Fermi momentum of nucleons in a projectile nucleus with a momentum per nucleon. The most valuable in this aspect are the events of dissociation, which are not accompanied by fragments of the target nuclei and generated mesons. They are called coherent dissociation or ``white'' stars. 

Despite the fact that the coherent dissociation $^{12}$C $\to$ 3$\alpha$ and $^{16}$O $\to$ 4$\alpha$ is only 1–2\%, the targeted search performed by transverse scanning, it is possible to investigate 310 3$\alpha$ and 641 4$\alpha$ ``white'' stars \cite{BelagaPAN:1995,AndreevaPAN:1996} by using the invariant mass method and establish the contributions of 3$\alpha$-decays of the Hoyle state (HS) \cite{ArtemRM:2018,ArtemFewBP:2020} in both cases. In general, the invariant mass $Q$ = $M^{*}$ - $M$ is given by the sum $M^{*2}$ = $\Sigma(P_i\cdot P_k)$, where $P_{i,k}$ are 4-momenta of fragments, and $M$ is their mass. To calculate the invariant masses of 2$\alpha$-pairs $Q_{2\alpha}$ and 3$\alpha$-triplets $Q_{3\alpha}$ in the approximation of conservation of momentum per nucleon by $\alpha$-particles of the primary nucleus, only measurements of their emission angles are used. The correspondence between He - $^{4}$He and H - $^{1}$H is assumed, since in the case of extremely narrow decays of $^{8}$Be and $^{9}$B, the measured contributions of $^{3}$He and $^{2}$H are small. The initial portions of the event distributions over the variables $Q_{2\alpha}$ and $Q_{3\alpha}$ contain peaks corresponding to $^{8}$Be and HS for both $^{12}$C and $^{16}$O. The selection $Q_{2\alpha}$($^{8}$Be) $\leq$ 0.2 MeV and $Q_{3\alpha}$(HS) $\leq$ 0.7 MeV is possible since the decay energy values are noticeably lower than the nearest excitations. Their application gives a contribution of $^{8}$Be (HS) 45 $\pm$ 4\% (11 $\pm$ 3\%) for $^{12}$C and 62 $\pm$ 3\% (22 $\pm$ 2\%) - for $^{16}$O. 

The invariant approach helps to identify the decays  $^8$Be, $^9$B, and HS, including the cascade ones among the relativistic fragments independently on the initial collision energy. It becomes possible to establish a connection with the low energy studies \cite{Borderie:2016,Barbui:2018, Bishop:2019, Kokalova:2020}. The effect of relativistic collimation can be used not only to investigate the generation of $^8$Be, $^9$B and HS, but also search for unstable states of increasing complexity decaying through them \cite{ZarubinLN:2013, ArtemPPN:2017, ArtemPAN:2017}. The feasibility of this approach with other methods of high-energy physics has not been demonstrated yet. 

Earlier, the contribution of $^8$Be and $^9$B decays to the dissociation of few light, medium (Ne, Si) and heavy (Au) nuclei were estimated in a similar way (review \cite{ArtemEPJ:2020}). Each of these unstable states has extremely low decay energy and lifetime (inversely proportional to the widths) and is several orders higher than the characteristic time of generating reactions. They are predicted to be unusually large in size (example in \cite{Schuck:2017}). One can assume the presence of these unstable states as virtual components in parent nuclei, which manifest themselves in relativistic fragmentation. However, maintaining this universality with the increase of the mass number of nuclei under study seems to be more and more problematic. The alternative consists in the $^{8}$Be formation during the final state interaction of the produced $\alpha$-particles and subsequent pick-up of accompanying $\alpha$-particles and nucleons with the necessary $\gamma$-quanta emission. The consequence of this scenario would be the increase in the $^{8}$Be yield with the multiplicity of $\alpha$-particles in the event $n_{\alpha}$ and, probably, $^{9}$B and HS decaying through $^{8}$Be. The purpose of this study is to identify the relationship between the formation of unstable states and accompanying multiplicity $n_{\alpha}$. 

The smaller the difference between the charges and mass numbers of the parent nucleus and the reconstructed unstable state, the easier they are identified (for example, $^9$Be $\to$ $^8$Be and $^{10}$C $\to$ $^9$B \cite{ArtemEPJ:2020}), since the distortions are minimized in determining the fragment emission angles which tend to increase in the transition from track to track. In addition, the combinatorial background from the accompanying multiplicity is minimized in the studied region of invariant masses. However, the above limitation related with multiplicity slows down testing the universality and correlations in the unstable state production as well as search for more complex states of this kind. NTE layers exposed to heavy nuclei radically expand the multiplicity of the studied fragments that requires to study identification conditions with increasing $n_{\alpha}$ in practice.

The primary track tracing in NTE allows one to find interactions without sampling with a different number of relativistic fragments of He and H. The data obtained by using this approach have traced the contribution of the unstable states and provided an opportunity for applying the advanced transverse scanning method to get more statistics and complex states. Although the multiple channel statistics turns out to be radically lower but its evolution can be seen. Below the article gives the overview of the measurements gathered by the emulsion collaboration at the JINR Synchrophasotron in the 80s and EMU collaboration at the AGS (BNL) and SPS (CERN) synchrotrons in the 90s on the fragmentation of relativistic nuclei $^{16}$O, $^{22}$Ne, $^{28}$Si and $^{197}$Au \cite{AndreevaSovJNP:1988, NaghyJPG:1988, AdamovichPRC:1989, AdamovichZPA:1995, AdamovichEPJA:1999}. Photos and videos of characteristic interactions are available \cite{ZarubinLN:2013, BecquerelWeb}.

These data have preserved their uniqueness in terms of relativistic nuclear fragmentation being large-scaled and uniformed. A fundamental value of the conclusion is presented concerning the limiting fragmentation regime in the widest possible range of nuclei and primary energy values expressed by the invariability of the charge composition of fragments and scale-invariant behavior of their spectra. At the same time the fine effects associated with angular correlations within the ensembles of fragments remained unexplored despite the diversity of the results obtained. In addition to the actual interest in the topic, they require a targeted build-up of statistics in multiple channels. Owing to the use of NTE layers exposed at that time the statistics of the measured interactions $^{28}$Si $\to$ $n_{\alpha}$ ($\geq$ 3) has been started to be supplemented in the framework of our BECQUEREL experiment at JINR. All these measurements, uniformly represented in the variable of invariant mass, enable one to estimate the role of unstable states in multiple nuclear fragmentation and formulate the further tasks.

\section{\texorpdfstring{$^{16}$O}{16O} fragmentation }
\label{sec:1}
\begin{figure}
	\centering\includegraphics[width=13cm]{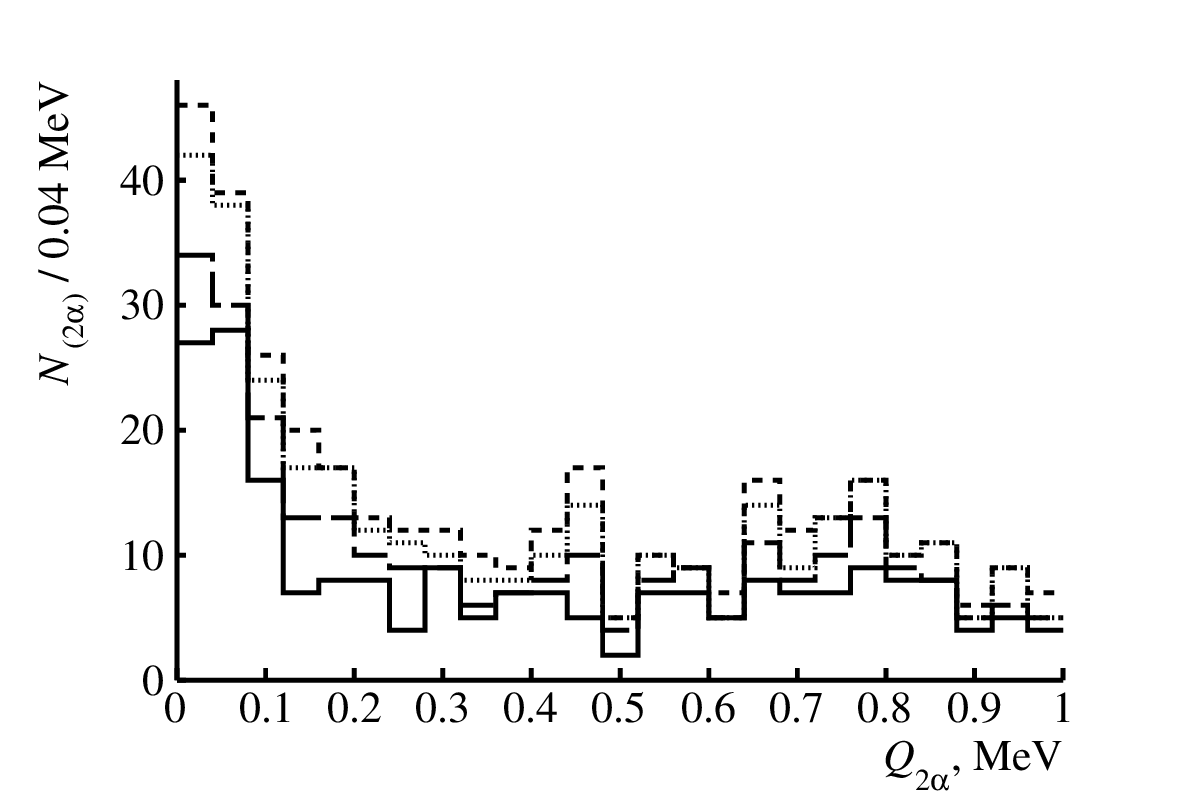}
	\caption{Distribution of 2$\alpha$-pairs $N_{(2\alpha)}$ over invariant mass $Q_{2\alpha}$ ($\leq$ 1 MeV) in fragmentation of 3.65 GeV/nucleon $^{16}$O nuclei (solid line); data for 15 (long dotted line), 60 (dotted line) and 200 (short-dotted line) GeV/nucleon $^{16}$O are added sequentially.}
	\label{fig:1}       
\end{figure}

There are measurements for inelastic interactions of $^{16}$O nuclei found while tracing primary tracks at four energy values including 2823 at 3.65 GeV/nucleon (JINR Synchrophasotron, 80s), 689 at 14.6 GeV/nucleon (BNL AGS, 90s), 885 at 60 GeV/nucleon and 801 at 200 GeV/~nucleon (CERN SPS). The distributions of all combinations of $\alpha$-pairs from these interactions $N_{(2\alpha)}$ over the invariant mass $Q_{2\alpha}$ $\leq$ 1 are stacked together in Fig.\ref{fig:1}. As in the case of high statistics $^{16}$O $\to$ 4$\alpha$ \cite{ArtemEPJ:2020}, the concentration of $\alpha$-pairs is observed at the beginning of the spectrum, and the condition $Q_{2\alpha}$($^8$Be) $\leq$ 0.2 MeV is accepted for selecting $^8$Be decay candidates.

\begin{table*}
	\caption{Statistics $N_{n\alpha}$($^8$Be) among $n_\alpha$ events of $^{16}$O dissociation; percentage of $N_{n\alpha}$($^8$Be) among $N_{n\alpha}$ is indicated.}
	\label{tab:1}
\centering \scalebox{0.92}{
		\begin{tabular}{cccccc}
			\hline
				\noalign{\smallskip}
			$n_{\alpha}$ 
			&\begin{tabular}{cc}
				3.65 GeV/nucleon \\
				$N_{n\alpha}$($^8$Be)/$N_{n\alpha}$(\%) 
			\end{tabular} 
			&\begin{tabular}{cc}
				15 GeV/nucleon \\
				$N_{n\alpha}$($^8$Be)/$N_{n\alpha}$(\%) 
			\end{tabular} 
			&\begin{tabular}{cc}
				60 GeV/nucleon \\
				$N_{n\alpha}$($^8$Be)/$N_{n\alpha}$(\%) 
			\end{tabular} 
			& \begin{tabular}{cc}
				200 GeV/nucleon \\
				$N_{n\alpha}$($^8$Be)/$N_{n\alpha}$(\%) 
			\end{tabular} 
			& \begin{tabular}{cc}
				All \\
				$N_{n\alpha}$($^8$Be)/$N_{n\alpha}$(\%) 
			\end{tabular}  \\
			\noalign{\smallskip}\hline\noalign{\smallskip}
			2  & 32/390 (8 $\pm$ 2) & 6/95 (6 $\pm$ 3) & 9/97 (9 $\pm$ 3) & 3/56 (5 $\pm$ 3) & 50/638 (8 $\pm$ 1) \\ \noalign{\smallskip}
			3  &  40/176 (23 $\pm$ 4) & 13/51 (26 $\pm$ 8) & 12/64 (19 $\pm$ 6) & 8/29 (28 $\pm$ 11) & 73/320 (23 $\pm$ 3) \\ \noalign{\smallskip}
			4  &  13/28 (46 $\pm$ 15) &	1/4 (25) &	2/2 (100) &	0/1 (0) & 16/35 (46 $\pm$ 14)  \\
			\noalign{\smallskip}\hline
		\end{tabular}
}
\end{table*}

Table \ref{tab:1} shows the number of events $N_{n\alpha}$($^8$Be) containing at least one $^8$Be decay candidate satisfying the condition $Q_{2\alpha}$($^8$Be) $\leq$ 0.2 MeV, among the events $N_{n\alpha}$ with the relativistic $\alpha$-particle multiplicity $n_{\alpha}$. In the covered initial energy range the distributions $N_{n\alpha}$ and $N_{n\alpha}$($^8$Be) have shown similarities, which corresponds to the concept of the limiting nuclear fragmentation regime. As $n_{\alpha}$ increases, the fraction of events with $^8$Be decays increases. The invariability of the composition of relativistic fragmentation from the initial energy gives grounds to summarize the statistics confirming the contribution of $^8$Be which grows with $n_{\alpha}$ (right column of Table \ref{tab:1}). 13 4$\alpha$-events are ``white'' stars, and 6 of them contain $^8$Be decays. This number relates the ``white'' 4$\alpha$-star statistics mentioned above with the other $^{16}$O dissociation channels.

\begin{table*}
	\caption{Statistics of $N_{n\alpha mp}$($^9$B) and $N_{n\alpha mp}$($^8$Be) decays in the fragmentation channels $N_{n\alpha mp}$($^8$Be) of $^{16}$O, $^{22}$Ne and $^{28}$Si nuclei with a multiplicity of $\alpha$-particles $n_\alpha$ and protons $m_p$.}
	\label{tab:2}
	\begin{center}
		\begin{tabular}{cccccc}
			\hline\noalign{\smallskip}
			$^{16}$O & $^{16}$O & $^{22}$Ne & $^{22}$Ne & $^{28}$Si & $^{28}$Si
			\\
			$N_{n\alpha mp}$ & $\frac{N_{n\alpha mp}(^9\textrm{B})}{N_{n\alpha mp}(^8\textrm{Be})}$ (\%) 

             & $N_{n\alpha mp}$ &	$\frac{N_{n\alpha mp}(^9\textrm{B})}{N_{n\alpha mp}(^8\textrm{Be})}$ (\%) &
			$N_{n\alpha mp}$ &	$\frac{N_{n\alpha mp}(^9\textrm{B})}{N_{n\alpha mp}(^8\textrm{Be})}$ (\%)\\
			\noalign{\smallskip}\hline\noalign{\smallskip}
			338 2$\alpha$ + (1-4)$p$ & 9/26 (35 $\pm$ 14) & 429 2$\alpha$ + (1-6)$p$ & 8/25 (32 $\pm$ 13) & 184 2$\alpha$ + $mp$ & 2/8 (25 $\pm$ 20)
			 \\ \noalign{\smallskip}
			 
			131 3$\alpha$ + (1,2)$p$ & 12/31 (39 $\pm$ 13) & 203 3$\alpha$ + (1-4)$p$ & 8/39 (21 $\pm$ 8) & 320 3$\alpha$ + $mp$ & 8/47 (17 $\pm$ 7) \\ \noalign{\smallskip}
			
			- & - &	58 4$\alpha$ + (1,2)$p$ & 5/20 (25 $\pm$ 12) & 168 4$\alpha$ + $mp$ & 9/55 (16 $\pm$ 6) \\  \noalign{\smallskip}
			
			- & - & - & - & 62 5$\alpha$ + $mp$ & 3/24 (13 $\pm$ 8) \\  \noalign{\smallskip}
			
			- & - & - & - & 7 6$\alpha$ + $mp$ & 0/5 \\
			
			\noalign{\smallskip}\hline

		\end{tabular}
	\end{center}
\end{table*}

\begin{figure}
	\centering\includegraphics[width=13cm]{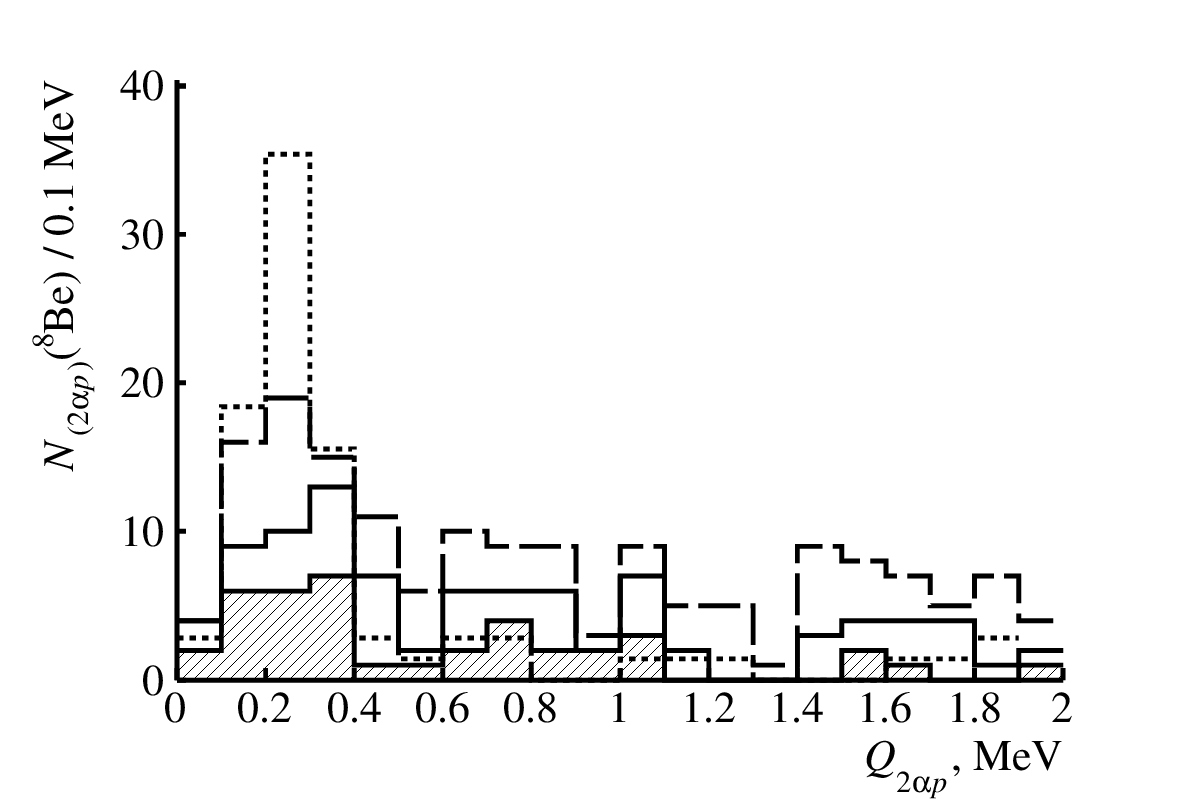}
	\caption{Distribution of 2$\alpha p$ triples $N_{(2\alpha p)}$($^8$Be) over invariant mass $Q_{2\alpha p}$ $\leq $ 2 MeV in 3.65 GeV/nucleon $^{16}$O (shaded), 3.22 GeV/nucleon $^{22}$Ne fragmentation (added by the solid line) and 15 GeV/nucleon $^{28}$Si (added by the dashed line). Dots mark $N_{2\alpha p}$($^8$Be) distribution in coherent dissociation $^{10}$C $\to$ 2$\alpha$2$p$ (normalized to $^{16}$O, $^{22}$Ne and $^{28}$Si statistics).}
	\label{fig:2}
\end{figure}

Only the $N_{n\alpha}$($^8$Be) statistics at 3.65 GeV/nucleon corresponds to the level expected for the $^9$B and HS decays. The number of 2$\alpha p$ triples $N_{2\alpha p}$($^8$Be) under the condition $Q_{2\alpha}$($^8$Be) $\leq$ 0.2 MeV noticeably increases at the beginning of the spectrum at $Q_{2\alpha p}$($^9$B) $\leq$ 0.5 MeV (Fig. \ref{fig:2}). This criterion has been taken for the $^9$B decays $N_{2\alpha p}$($^9$B). It coincides with the extraction of 54 $^9$B decays in the most convenient of coherent dissociation $^{10}$C $\to$ 2$\alpha$2$p$ at 1.2 GeV/nucleon (Fig. \ref{fig:2}) \cite{ArtemEPJ:2020}.

In the channels $n_{\alpha}$ with the multiplicity of protons $mp$, on going from $n_{\alpha}$ = 2 to 3, the number $N_{n\alpha mp}$($^9$B) increases relatively to $N_{n\alpha mp}$($^9$B) and  proportionally to $N_{n\alpha mp}$($^8$Be) (Table \ref{tab:2}). One HS decay is identified at $n_{\alpha}$ = 3 and 5 - at $n_{\alpha}$ = 4. In the latter case, $N_{4\alpha}$(HS)/$N_{n\alpha}$($^8$Be) = 0.4 $\pm$ 0.2 does not contradict the result for ``white'' 4$\alpha$ stars.

\section{Fragmentation of \texorpdfstring{$^{16}$O}{16O} on protons}
\label{sec:2}
The accepted approximations can be verified using the data obtained in exposure to 2.4 GeV/nucleon $^{16}$O nuclei of the JINR 1-meter hydrogen bubble chamber (VPK-100), placed in the magnetic field \cite{GlagolevEPJA:2001}. The dataset includes full solid angle measurements of the momentum vectors of the $^{16}$O + $p$ reaction products in 11104 collisions of all kinds. In this case, there is also the concentration of $\alpha$-pairs in the initial part of the angle distribution $\Theta_{2\alpha}$, corresponding to $^8$Be decays \cite{GlagolevEPJA:2001}. As noted \cite{ArtemEPJ:2020}, when momenta of relativistic He fragments reconstructed with insufficient accuracy are used in the $Q_{2\alpha}$ calculation, the $^8$Be signal practically disappears. There is still an opportunity of momentum fixing (as in the NTE case) and using the measured values while normalizing to the value of the initial momentum per nucleon to identify of He and H isotopes.

\begin{figure}
	\centering\includegraphics[width=13cm]{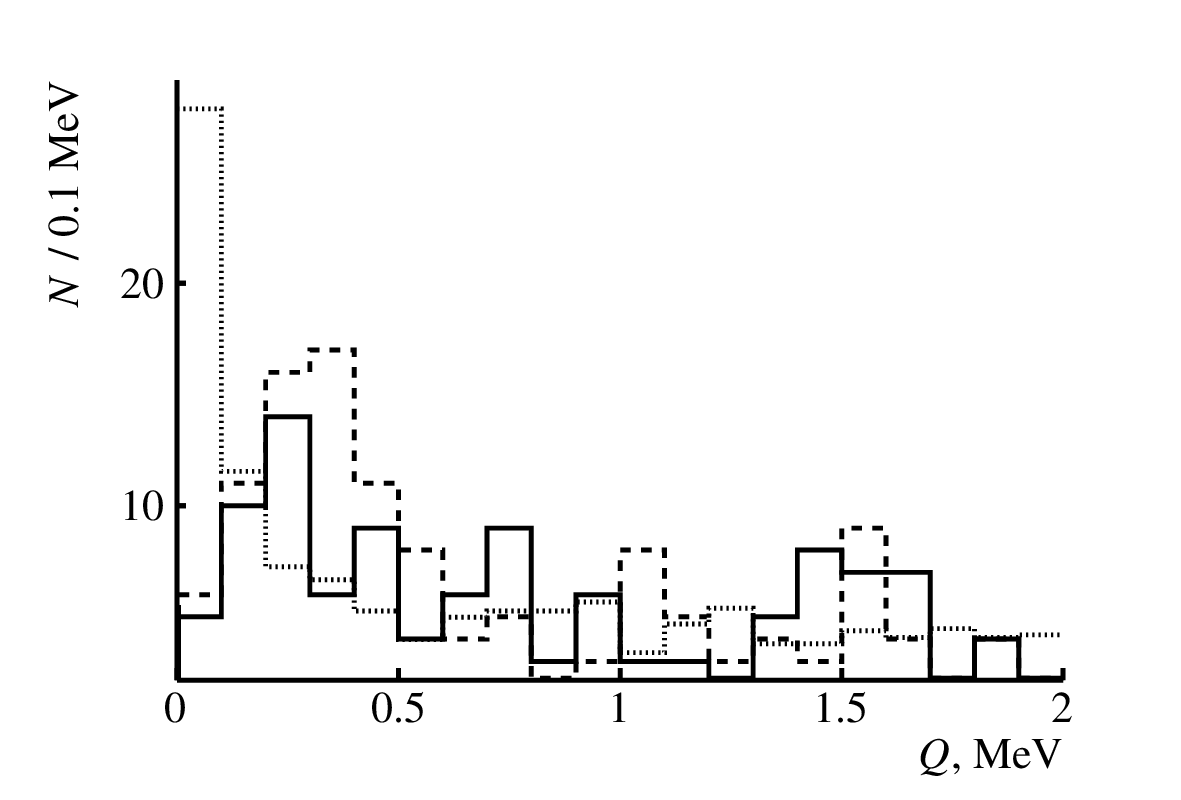}
	\caption{Distribution of 2.4 GeV/nucleon $^{16}$O fragmentation events on protons over invariant masses of all 2$\alpha$-pairs $Q_{2\alpha}$ (dots), 2$\alpha p$-triplets $Q_{2\alpha p}$ (dashed line) and 3$\alpha$-triplets $Q_{3\alpha}$ (solid).}
	\label{fig:3}       
\end{figure}

The invariant mass distributions of all $\alpha$-pairs $Q_{2\alpha}$, 2$\alpha p$-triplets $Q_{2\alpha p}$, and 3$\alpha$-triplets $Q_{3\alpha}$ calculated from the angles determined in the VPK-100, are superimposed in Fig. \ref{fig:3}. In the presented range, the $Q_{2\alpha}$ distribution is normalized to the $Q_{2\alpha p}$ statistics with a decreasing factor of 25. Directly depending on $\Theta_{2\alpha}$, the $Q_{2\alpha}$ variant with fixed momenta demonstrates the signal of $^8$Be. According to the measured momenta of fragments, the condition $Q_{2\alpha}$($^8$Be) $\leq$ 2 MeV removes the 3He contribution, and the contribution of protons is 90\% among H.

The distribution $Q_{2\alpha p}$ shown in Fig. \ref{fig:3} indicates 60 $^9$B decays to $Q_{2\alpha p}$($^9$B) $\leq$ 0.5 MeV with the condition $Q_{2\alpha}$($^8$Be) $\leq$ 0.2 MeV and selection of protons. The average value $\left\langle Q_{2\alpha p}\right\rangle$ (RMS) = 271 $\pm$ 15 (120) keV corresponds to the NTE result \cite{ArtemRM:2018}. Similarly, in the distribution $Q_{3\alpha}$(HS) $\leq$ 0.7 MeV under the condition $Q_{2\alpha}$($^8$Be) $\leq$ 0.2 MeV there are 47 HS decays identified in the peak having $\left\langle Q_{3\alpha}\right\rangle$ (RMS) = 322 $\pm$ 25 (180) keV \cite{ArtemRM:2018}. In the statistics, there are four events with the 2$^8$Be formation and one event - with coincident candidates $^9$B and HS.

Table \ref{tab:3} shows the data reflecting the change in contributions from unstable state decays in the events with the $\alpha$-particle multiplicity $n_{\alpha}$ (in this case, identified $^4$He nuclei). With increasing $n_{\alpha}$, the $^8$Be detecting probability increases rapidly. The increase in $n_{\alpha}$ results in relative decrease of $N_{n\alpha}$($^9$B), which can be explained by the decrease in the number of protons available for the $^9$B formation. On the contrary, $N_{n\alpha}$(HS) increases due to the increase of the number of $\alpha$-particles available for the HS formation. In the coherent dissociation $^{16}$O $\to$ 4$\alpha$, the fraction of HS decays relatively to $^8$Be is 35 $\pm$ 4\%, which does not contradict to the value for $n_{\alpha}$ = 4 in the generally more severe $^{16}$O + $p$ interaction (Table \ref{tab:3}). These facts have indicated the universality of the appearance of $^8$Be and HS.

\begin{table}
	\caption{Statistics of events containing at least one $^8$Be decay candidate $N_{n\alpha}$ ($^8$Be), $^9$B, or HS under the condition $Q_{2\alpha}$($^8$Be) $\leq$ 0.2 MeV among the $N_{n\alpha}$ events of fragmentation of $^{16}$O nucleus fragmentation on protons with multiplicity $n_{\alpha}$.}
	\label{tab:3}
\centering \scalebox{0.9}{
		\begin{tabular}{cccc}
			\hline
			\noalign{\smallskip}
			$n_{\alpha}$ 
			&\begin{tabular}{cc}
				$N_{n\alpha}$($^8$Be)/$N_{n\alpha}$ \\
				(\% $N_{n\alpha}$) 
			\end{tabular}
			&\begin{tabular}{cc}
				$N_{n\alpha}$($^9$B)\\
				(\% $N_{n\alpha}$($^8$Be))
			\end{tabular}
			&\begin{tabular}{cc}
				$N_{n\alpha}$(HS)\\
				(\% $N_{n\alpha}$($^8$Be))
			\end{tabular}\\
			\noalign{\smallskip}\hline\noalign{\smallskip}
			2 & 111/981 (11 $\pm$ 1) & 29 (26 $\pm$ 6) & - \\ \noalign{\smallskip}
			3 & 203/522 (39 $\pm$ 3) & 31 (15 $\pm$ 3) & 36 (18 $\pm$ 3) \\ \noalign{\smallskip}
			4 & 27/56 (48 $\pm$ 11) & - & 11 (41 $\pm$ 15) \\ \noalign{\smallskip}
			\noalign{\smallskip}\hline
		\end{tabular}
}
\end{table}

\section{\texorpdfstring{$^{22}$N\lowercase{e}}{22Ne} and \texorpdfstring{$^{28}$S\lowercase{i}}{28Si} fragmentation}
\label{sec:3}
\begin{figure}
	\centering\includegraphics[width=13cm]{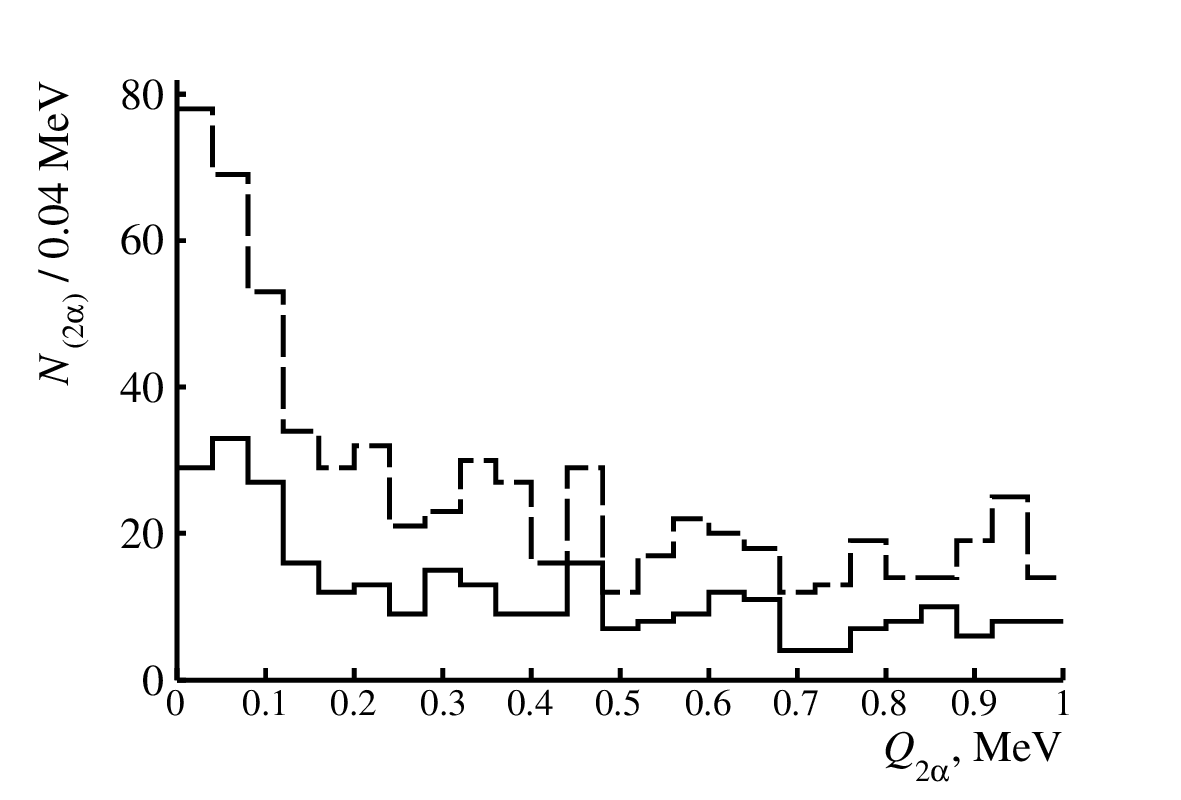}
	\caption{Distribution of 2$\alpha$-pairs $N_{(2\alpha)}$ over invariant mass $Q_{2\alpha}$ ($\leq$ 1 MeV) in fragmentation of 3.22 GeV/nucleon $^{22}$Ne (solid line) and 14.6 GeV/nucleon $^{28}$Si nuclei (added by the dotted line).}
	\label{fig:4}       
\end{figure}

The measurements carried out in NTE layers exposed to $^{22}$Ne nuclei at 3.22 GeV/nucleon (4308 events, JINR Synchrophasotron) and $^{28}$Si at 14.6 GeV/nucleon (1093 events, BNL AGS) further expanded the $n_{\alpha}$ range. In both cases, no change in the condition $Q_{2\alpha}$($^8$Be) $\leq$ 0.2 MeV is required (Fig. \ref{fig:4}). The $N_{n\alpha}$ and $N_{n\alpha}$($^8$Be) statistics are presented in Table \ref{tab:4}. Recently, the $^{28}$Si statistics $n_\alpha$ $\geq$ 3 has been tripled by transverse scanning (Table \ref{tab:4}). In these cases, the $^8$Be contribution also increases with the $n_\alpha$ multiplicity. \\

\begin{figure}
	\centering\includegraphics[width=13cm]{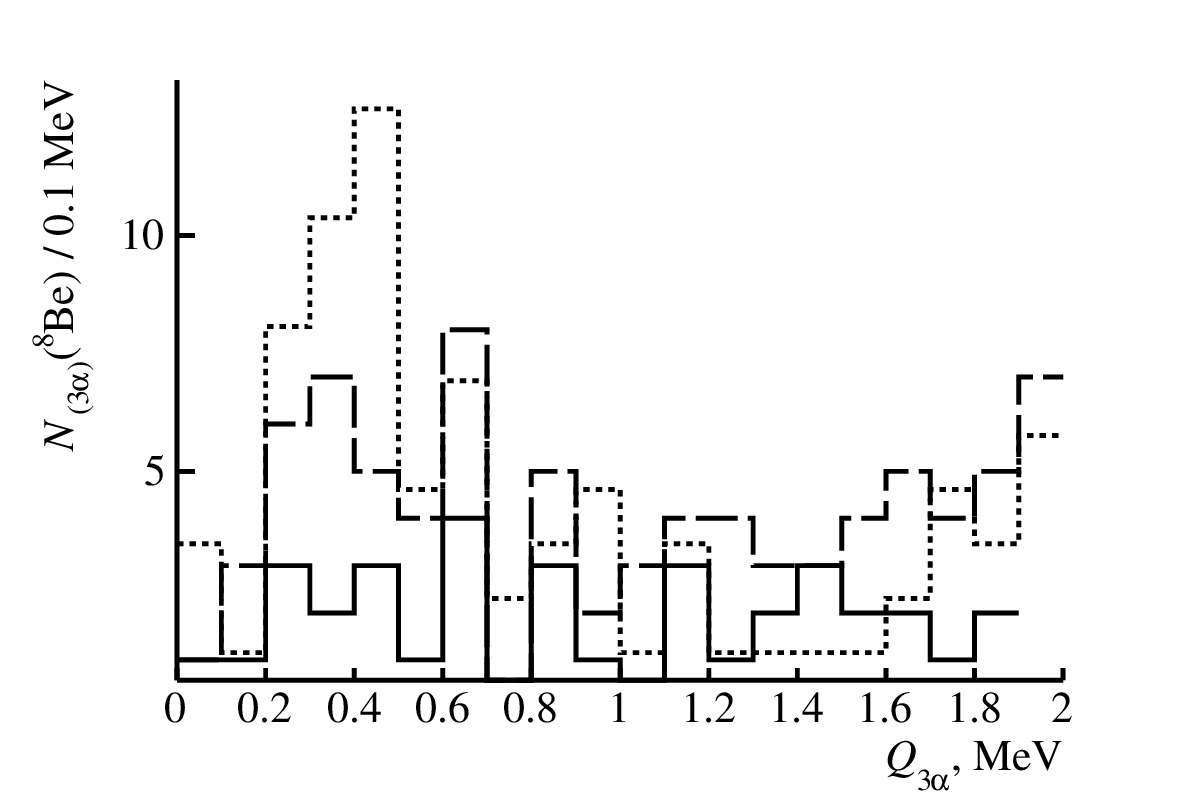}
	\caption{Distribution of 3$\alpha$-triplets $N_{(3\alpha)}$($^8$Be) over invariant mass $Q_{2\alpha p}$ ($\leq$ 2 MeV) in 3.22 GeV/nucleon $^{22}$Ne fragmentation (solid line) and $^{28}$Si at 14.6 GeV/nucleon (dashed line). Dots mark distribution $N_{(3\alpha)}$($^8$Be) in dissociation $^{12}$C $\to$ 3$\alpha$ normalized to $^{22}$Ne and $^{28}$Si statistics.}
	\label{fig:5}       
\end{figure}

The $^8$Be nuclei can appear both: directly in fragmentation and via $^9$B and HS decays. Only the $^{22}$Ne statistics allowed one to estimate the contributions of $^9$B and HS based on the invariant masses $Q_{2\alpha p}$ and $Q_{3\alpha}$. The distribution $N_{(2\alpha p)}$($^8$Be), added in Fig. \ref{fig:2} for the $^{22}$Ne case, contains 2$\alpha p$ triples $Q_{2\alpha p}$($^9$B) $\leq$ 0.5 MeV. The $N_{(3\alpha)}$($^8$Be) distribution shown in Fig. \ref{fig:5}, contains 3$\alpha$ triples $Q_{3\alpha}$(HS) $\leq$ 0.7 MeV. Previously, this condition was used to establish 41 candidates for HS decays in the dissociation $^{12}$C $\to$ 3$\alpha$ from the $N_{(3\alpha)}$($^8$Be) distribution in Fig. \ref{fig:4}. Similarly to the case of $^{16}$O, on going from $n_\alpha$ = 2 to 4 for $^9$B, $N_{n\alpha mp}$($^9$B) relative to $N_{n\alpha mp}$ increases (Table~\ref{tab:2}). Transition from $n_\alpha$ = 3 to 4 indicates a noticeable increase in HS, showing the analogy with the $^{16}$O data as well (Table~\ref{tab:5}).

\begin{table}
	\caption{Statistics of events $N_{n\alpha}$($^8$Be) among $N_{n\alpha}$ events in dissociation of $^{22}$Ne and $^{28}$Si nuclei; the percentage of $N_{n\alpha}$($^8$Be) among  $N_{n\alpha}$ is indicated.}
	\label{tab:4}
	\begin{center}
		\begin{tabular}{ccc}
			\hline
			\noalign{\smallskip}
			$n_{\alpha}$ 
			&\begin{tabular}{cc}
				$^{22}$Ne 3.22 GeV/nucleon \\
				$N_{n\alpha}$($^8$Be)/$N_{n\alpha}$ (\%)
			\end{tabular}
			&\begin{tabular}{cc}
				$^{28}$Si 15 GeV/nucleon \\
				$N_{n\alpha}$($^8$Be)/$N_{n\alpha}$ (\%)
			\end{tabular}\\
			\noalign{\smallskip}\hline\noalign{\smallskip}
			2 &	30/528 (6 $\pm$ 1) & 5/164 (3 $\pm$ 2)  \\ \noalign{\smallskip}
			3 & 45/243 (19 $\pm$ 3) & 47/320 (15 $\pm$ 2)  \\ \noalign{\smallskip}
			4 & 25/80 (31 $\pm$ 6) & 55/168 (33 $\pm$ 5)  \\ \noalign{\smallskip}
			5 & 6/10 (60 $\pm$ 31) & 24/62 (39 $\pm$ 10)  \\ \noalign{\smallskip}
			6 & - & 5/7 (72 $\pm$ 42) \\ \noalign{\smallskip}
			\noalign{\smallskip}\hline
		\end{tabular}
	\end{center}
\end{table}

\begin{table*}
	\caption{Statistics of events $N_{n\alpha}$(HS) among events $N_{n\alpha}$ in dissociation of $^{22}$Ne nuclei and $^{28}$Si; the percentage of $N_{n\alpha}$($^8$Be) among $N_{n\alpha}$ is indicated.}
	\label{tab:5}
	\begin{center}
		\begin{tabular}{ccccc}
			\hline\noalign{\smallskip}
			 & $^{22}$Ne & $^{22}$Ne & $^{28}$Si & $^{28}$Si \\
			$n_{\alpha}$ & $N_{n\alpha}$(HS)/$N_{n\alpha}$ (\%$N_{n\alpha}$) & $N$(HS)/$N_{n\alpha}$($^8$Be), \% & 
			$N_{n\alpha}$(HS)/$N_{n\alpha}$ (\%$N_{n\alpha}$) & $N$(HS)/$N_{n\alpha}$($^8$Be), \% \\
			\noalign{\smallskip}\hline\noalign{\smallskip}
			3 &	3/243 (1.2 $\pm$ 0.6) & 7 $\pm$ 4 &	4/320 (1.2 $\pm$ 0.6) & 4/47 (9 $\pm$ 5) \\ \noalign{\smallskip}
			4 & 10/80 (13 $\pm$ 5) & 40 $\pm$ 15 &	7/168 (4 $\pm$ 2) & 7/55 (13 $\pm$ 5) \\ \noalign{\smallskip}
			5 & 1/10 & 17 &	7/62 (11 $\pm$ 5) & 7/24 (29 $\pm$ 12) \\
			6 & - & - & 1/7 (14) & 1/5 (20) \\
			\noalign{\smallskip}\hline
		\end{tabular}
	\end{center}
\end{table*}

\begin{table*}
	\caption{Statistics of events containing at least one $^8$Be, $^9$B or HS decay, or at least two $^8$Be provided $Q_{2\alpha}$ ($^8$Be) $\leq$ 0.4 MeV among the events $N_{n\alpha}$ of $^{197}$Au fragmentation with multiplicity $n_\alpha$; the total statistics of the channels $n_\alpha$ $\geq$ 11 is given in italics.}
	\label{tab:6}
	\begin{center}
		\begin{tabular}{ccccc}
			\hline\noalign{\smallskip}
			$n_{\alpha}$ 
			&\begin{tabular}{cc}
				$N_{n\alpha}$($^8$Be)/$N_{n\alpha}$ \\
				(\% $N_{n\alpha}$)
			\end{tabular}
			&\begin{tabular}{cc}
				$N_{n\alpha}$($^9$B) \\
				(\% $N_{n\alpha}$($^8$Be))
			\end{tabular}
			&\begin{tabular}{cc}
				$N_{n\alpha}$(HS) \\
				(\% $N_{n\alpha}$($^8$Be))
			\end{tabular}
			&\begin{tabular}{cc}
				$N_{n\alpha}$(2$^8$Be) \\
				(\% $N_{n\alpha}$($^8$Be))
			\end{tabular}\\
			\noalign{\smallskip}\hline\noalign{\smallskip}
			2 & 3/133 (2 $\pm$ 1) &	- &	- &	- \\ \noalign{\smallskip}
			3 &	14/162 (9 $\pm$ 3) & 1 (7) & - & - \\ \noalign{\smallskip}
			4 & 25/161 (16 $\pm$ 4) & 7 (28 $\pm$ 12) & 2 (8 $\pm$ 6) & - \\ \noalign{\smallskip}
			5 & 23/135 (17 $\pm$ 4) & 5 (22 $\pm$ 11) &	- &	1 (4) \\ \noalign{\smallskip}
			6 & 31/101 (31 $\pm$ 7) & 9 (29 $\pm$ 11) & 2 (6 $\pm$ 4) & - \\  \noalign{\smallskip}
			7 & 31/90 (34 $\pm$ 7) & 6 (19 $\pm$ 9) & 2 (6 $\pm$ 4) & 3 (10 $\pm$ 6) \\ \noalign{\smallskip}
			8 & 32/71 (45 $\pm$ 10) & 8 (25 $\pm$ 10) & 2 (6 $\pm$ 4) & 2 (7 $\pm$ 5) \\  \noalign{\smallskip}
			9 & 29/54 (54 $\pm$ 13) & 9 (31 $\pm$ 12) & 3 (10 $\pm$ 6) & 5(17 $\pm$ 8) \\  \noalign{\smallskip}
			10 & 22/39 (56 $\pm$ 15) & 4 (18 $\pm$ 10) & - & 5(23 $\pm$ 12) \\  \noalign{\smallskip}
			11 &\begin{tabular}{cc}
				10/15 (67 $\pm$ 27) \\
				$\textit{19/30 (63 $\pm$ 19)}$
			\end{tabular}
			&\begin{tabular}{cc}
				3 (30 $\pm$ 20) \\
				$\textit{7 (37 $\pm$ 16)}$
			\end{tabular}
			&\begin{tabular}{cc}
				1 (10) \\
				$\textit{2(11 $\pm$ 8)}$
			\end{tabular}
			&\begin{tabular}{cc}
				2(20 $\pm$ 16) \\
				$\textit{6 (32 $\pm$ 15)}$
			\end{tabular}\\  \noalign{\smallskip}
			12 & 2/5 & 1 & - & 1 \\ \noalign{\smallskip}
			13 & 2/4 & 1 & - & 1 \\ \noalign{\smallskip}
			14 & 3/3 & 1 & - & 1 \\ \noalign{\smallskip}
			15 & 1/1 & - & - & - \\ \noalign{\smallskip}
			16 & 1/2 & 1 & 1 & 1 \\ \noalign{\smallskip}
			\noalign{\smallskip}\hline
		\end{tabular}
		\end{center}
\end{table*}

\section{\texorpdfstring{$^{197}$A\lowercase{u}}{197Au} fragmentation}
\label{sec:4}
\begin{figure*}
\centering\includegraphics[width=15cm]{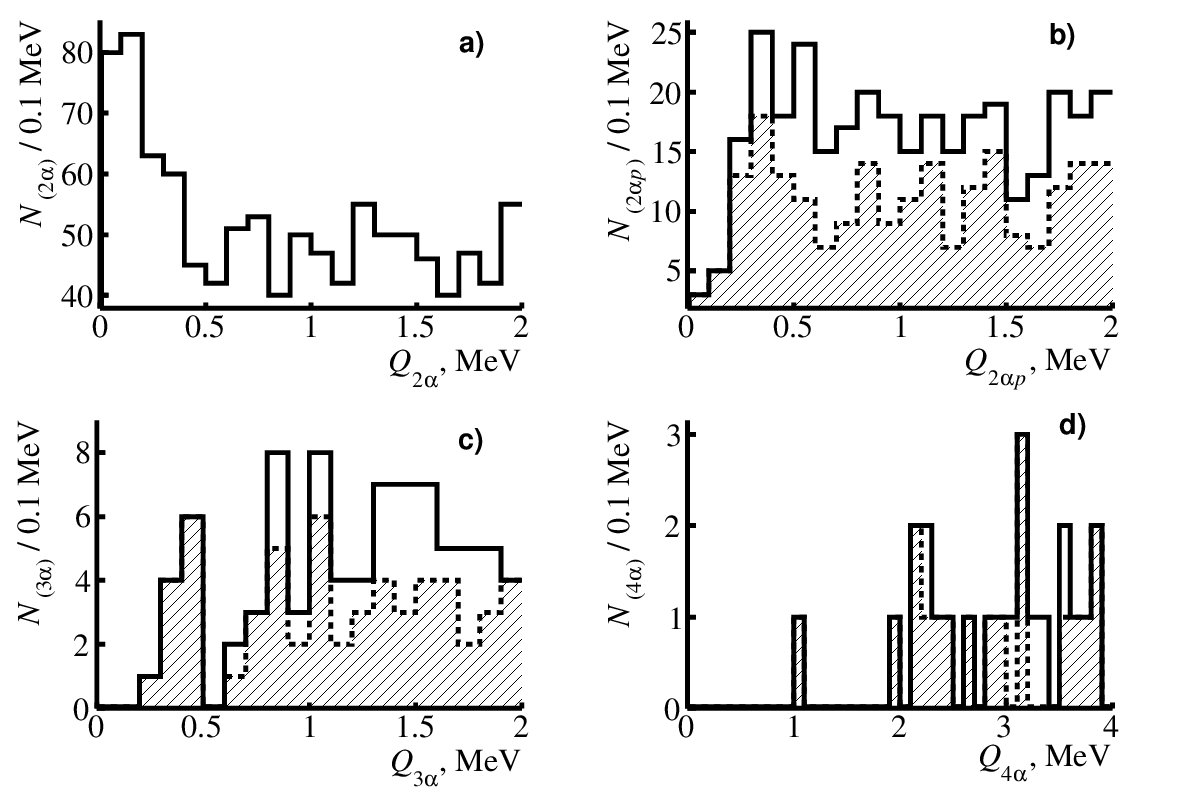}
	\caption{Distributions over invariant masses $Q$ of 2$\alpha$-pairs (a) in fragmentation of $^{197}$Au nuclei, as well as 2$\alpha p$-triplets (b), 3$\alpha$-triplets (c), and 4$\alpha$-quartets (d) in events with $^8$Be candidate selected according to $Q_{2\alpha}$($^8$Be) $\leq$ 0.4 MeV (solid) and $\leq$ 0.2 MeV (shaded).}
	\label{fig:6}       
\end{figure*}

There are similar measurements of 1316 interactions of $^{197}$Au nuclei at 10.7 GeV/nucleon (BNL AGS, 90s). For this dataset, Fig. \ref{fig:6}a shows the distribution of 2$\alpha$-pairs at small values of $Q_{2\alpha}$. Due to the deteriorated resolution, the $^8$Be region expands, which requires softening the selection of $Q_{2\alpha}$($^8$Be) $\leq$ 0.4 MeV to maintain the efficiency. Further, to check the $N_{n\alpha}$($^8$Be) correlation and minimize the background in $N_{n\alpha}$($^9$B) and $N_{n\alpha}$(HS), the condition $Q_{2\alpha}$($^8$Be) $\leq$ 0.2 MeV is also applied. Fig. \ref{fig:6} demonstrates the distributions of 2$\alpha p$-triplets, 3$\alpha$-triplets, and 4$\alpha$-quartets in the small $Q$ regions of the events where, according to these conditions, there is at least one candidate for $^8$Be decay. Note that the distributions (b) and (c) contain 2$\alpha p$ and 3$\alpha$ triples satisfying the conditions $Q_{2\alpha p}$($^9$B) $\leq$ 0.5 MeV and $Q_{3\alpha}$(HS) $\leq$ 0.7 MeV, respectively.

\begin{figure}
\centering\includegraphics[width=13cm]{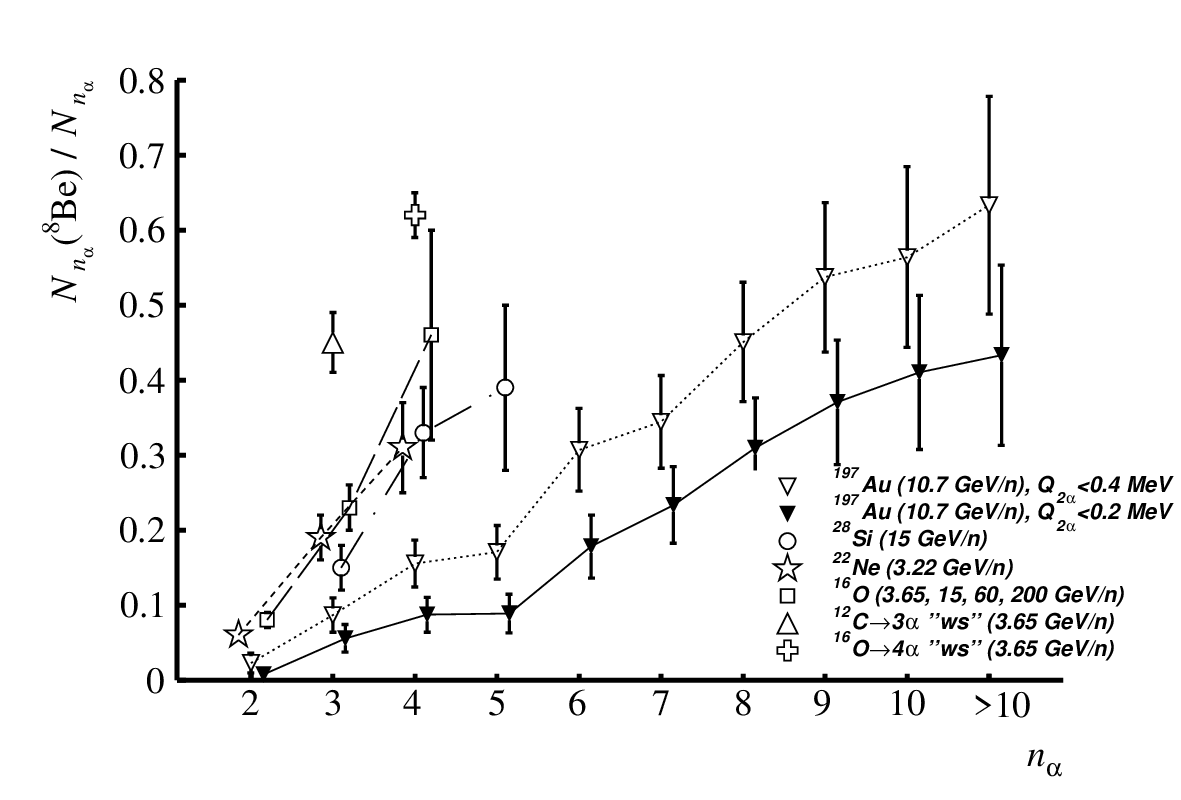}
	\caption{Dependence of relative contribution of $N_{n\alpha}$($^8$Be) decays to the statistics of $N_{n\alpha}$ events with multiplicity of $\alpha$-particles $n_\alpha$ in relativistic fragmentation of C, O, Ne, Si, and Au nuclei; ``white'' stars $^{12}$C $\to$ 3$\alpha$ and $^{16}$O $\to$ 4$\alpha$ are marked (WS); for convenience, the points are somewhat displaced around the values of $n_\alpha$ and are connected by lines.}
	\label{fig:7}       
\end{figure}

Statistics and relative yields of the unstable states for the $^{197}$Au nucleus are presented in Table \ref{tab:6} taking into account $Q_{2\alpha}$($^8$Be) $\leq$ 0.4 MeV. Channels $n_\alpha$ $\geq$ 11 are summed to reduce errors. The ratio of the number of events $N_{n\alpha}$ ($^8$Be) including at least one $^8$Be decay candidate to the statistics of the channel $N_{n\alpha}$, grows rapidly to $n_\alpha$ = 10 to about 0.5. This trend is preserved when the condition is tightened to $Q_{2\alpha}$($^8$Be) despite the decrease in statistics. All the discussed data on the ratio of the number of events $N_{n\alpha}$ ($^{8}$Be), including at least one candidate for $^{8}$Be decay, to the statistics of the $N_{n\alpha}$ channel, are combined in Fig. \ref{fig:7}.

The ratio of the number of events $N_{n\alpha}$($^9$B) and $N_{n\alpha}$(HS) to the statistics $N_{n\alpha}$($^8$Be) for the $^{197}$Au nucleus has not shown any noticeable change when multiplicity $n_\alpha$ alters (Table \ref{tab:6}). The statistics of the identified decays of $^8$Be pairs $N_{n\alpha}$(2$^8$Be) behaves the same way. In fact, these three ratios indicate the increase in $N_{n\alpha}$($^9$B), $N_{n\alpha}$(HS) and $N_{n\alpha}$(2$^8$Be) relatively to $N_{n\alpha}$. In these three cases, significant statistical errors allow one to characterize only the general trends. Summing the statistics on the multiplicity $n_\alpha$ and normalizing to the sum $N_{n\alpha}$($^8$Be) results in relative contributions $N_{n\alpha}$($^9$B), $N_{n\alpha}$(HS), and $N_{n\alpha}$(2$^8$Be) equal to 25 $\pm$ 4\%, 6 $\pm$ 2\%, and 10 $\pm$ 2\%, respectively.

The distribution over $Q_{4\alpha}$ (Fig.\ref{fig:6}d) indicates the near-threshold 4$\alpha$-quadruples where the decays of HS and 2$^8$Be are reconstructed under the condition $Q_{2\alpha}$($^8$Be) $\leq$ 0.2 MeV, including $Q_{4\alpha}$ = 1.0 (16$\alpha$, HS), 1.9 (11$\alpha$, HS), 2.1 (9$\alpha$, 2$^8$Be), 2.2 (5$\alpha$, 2$^8$Be), 2.4 (9$\alpha$, HS) MeV. The excited state 0$^+_6$ of the $^{16}$O nucleus mentioned above could decay along the chain $^{16}$O (0$^+_6$)  $\to$ $^{12}$C(0$^+_2$) $\to$ $^8$Be(0$^+$) $\to$ 2$\alpha$ or $^{16}$O(0$^+_6$) $\to$ 2$^8$Be(0$^+$) $\to$ 4$\alpha$. Investigations of this problem require a qualitatively different level of $n_\alpha$-ensemble statistics available in the transverse event search.

\section{Summary}
\label{sec:5}
The preserved and recently supplemented data on the relativistic fragmentation of $^{16}$O, $^{22}$Ne, $^{28}$Si, and $^{197}$Au nuclei in a nuclear track emulsion helped us to identify decays of $^{8}$Be, $^{9}$B nuclei and Hoyle state in the invariant mass distributions of 2$\alpha$-pairs, 2$\alpha p$- and 3$\alpha$-triplets. The determination of the invariant mass from the fragment emission angles in the velocity conservation approximation turns out to be an adequate approximation. Starting with the $^{16}$O fragmentation, the presented analysis has indicated a relative enhancement in the $^{8}$Be contribution while increasing in the number of relativistic $\alpha$-particles per event and remaining proportional contributions of HS and $^{9}$B. In the $^{197}$Au fragmentation, the tendency is traced up to at least 10 relativistic $\alpha$-particles per event. This observation has assumed the development of the theory of relativistic nucleus fragmentation taking into account the $\alpha$-particle interactions, that is characteristic for low-energy nuclear physics.

Taking to account all mentioned above it is necessary to increase the statistics of events with high multiplicity of $\alpha$-particles at high accuracy of measurements of the emission angles of relativistic He and H fragments. The analysis of the data on the $^{16}$O fragmentation in the hydrogen bubble chamber has confirmed the approximations and conclusions made. The application of this method would be efficient for light isotopes, including the radioactive ones. The feasibility of this approach in comparison with other methods in high energy physics has not been demonstrated yet. Therefore, the use of the flexible method of nuclear track emulsion retains a prospect to study the unstable states produced in a narrow cone of relativistic fragmentation by nuclei in the widest range of mass numbers. 

New opportunities are contained in existing layers exposed to 800-950 $A$ MeV $^{84}$Kr nuclei (SIS synchrotron, GSI, early 90s) which were already used for the reaction multiplicity survey \cite{KrasnovCzech:1996}. To limit the uncertainty associated with the deceleration of the beam nuclei, the analysis was performed on a small NTE section. In principle, the decrease in energy can be calculated and taken into account in the calculation of the invariant masses. Thus, the covered energy range and the viewed NTE area can be radically extended. This research is promising in the near future. It is important that reconstruction of $^{8}$Be and the Hoyle state in the presented approach has been successfully performed in the 400 MeV/nucleon $^{12}$C case \cite{ArtemRM:2018}.


\begin{thebibliography}{plain}
	
	\bibitem{Ajzenberg:1988}
	F. Ajzenberg-Selove, Nucl. Phys. A \textbf{490}, 1(1988); TUNL Nuclear Data Evaluation Project: \href{http://www.tunl.duke.edu/NuclData/}{http://www.tunl.duke.edu/NuclData/}.
	
	\bibitem{Beck:2012}
	C. Beck and P. Papka, Lect. Notes in Phys., \textbf{848}, Clusters in Nuclei, Volume 2, Springer Int. Publ., 289 (2012); \href{https://doi.org/10.1007/978-3-642-24707-1_6}{DOI: 10.1007/978-3-642-24707-1\_6.}
	
	\bibitem{Freer:2014}
	M. Freer and H. O. U. Fynbo, Progr. Part. Nucl. Phys. \textbf{78}, (2014);\\
	\href{https://doi.org/10.1016/j.ppnp.2014.06.001}{DOI: 10.1016/j.ppnp.2014.06.001.}	
	
	\bibitem{Smith:2020}
	R. Smith {et al}. Phys. Rev. C \textbf{101}, 2, 021302 (2020); \href{https://doi.org/10.1103/PhysRevC.101.021302}{DOI: 10.1103/PhysRevC.101.021302}
	
	\bibitem{Tohsaki:2001}
	A. Tohsaki, H. Horiuchi, P. Schuck, and G. Röpke  Phys. Rev.  Lett., \textbf{87}, 192501(2001); \\ \href{https://doi.org/10.1103/PhysRevLett.87.192501}{DOI: 10.1103/PhysRevLett.87.192501}
	
	\bibitem{Yamada:2004}
	T. Yamada and P. Schuck, Phys. Rev. C \textbf{69}, 024309 (2004); \\
	\href{https://doi.org/10.1103/PhysRevC.69.024309}{DOI: 10.1103/PhysRevC.69.024309.}
	
	\bibitem{Schuck:2017}
	A. Tohsaki, H. Horiuchi, P. Schuck and G. Röpke, Rev. Mod. Phys. \textbf{89}, 011002 (2017); \\ \href{https://doi.org/10.1103/RevModPhys.89.011002}{DOI: 10.1103/RevModPhys.89.011002.}
	
	\bibitem{Oertzen:2010}
	W. von Oertzen, Lect. Notes in Phys., \textbf{818}, Clusters in Nuclei, Volume 1. Springer Int. Publ., 109 (2010); \href{https://doi.org/10.1007/978-3-642-13899-7_3}{DOI: 10.1007/978-3-642-13899-7\_3.}
	
	\bibitem{Borderie:2016}
	B. Borderie \textit{et al}., Phys. Lett. B \textbf{755}, 475 (2016) ; \href{https://doi.org/10.1016/j.physletb.2016.02.061}{DOI: 10.1016/j.physletb.2016.02.061.}
	
	\bibitem{Barbui:2018}
	M. Barbui \textit{et al}., Phys. Rev. C \textbf{98}, 044601 (2018); \href{https://doi.org/10.1103/PhysRevC.98.044601}{DOI: 10.1103/PhysRevC.98.044601.}
	
	\bibitem{Bishop:2019}
	J. Bishop \textit{et al}., Phys. Rev. C \textbf{100}, 034320 (2019); \href{https://doi.org/10.1103/PhysRevC.100.034320}{DOI: 10.1103/PhysRevC.100.034320.}
	
	\bibitem{Kokalova:2020}
	R. Smith, J. Bishop, J. Hirst, Tz. Kokalova, C. Wheldon, Few Body Syst., \textbf{61} 2 (2020);\\ \href{https://doi.org/10.1007/s00601-020-1545-5}{DOI: 10.1007/s00601-020-1545-5.}
	
	\bibitem{ZarubinLN:2013}
	P.I. Zarubin, Lect. Notes in Phys., \textbf{875}, Clusters in Nuclei, Volume 3. Springer Int. Publ., 51 (2013); \href{https://doi.org/10.1007/978-3-319-01077-9_3}{DOI: 10.1007/978-3-319-01077-9\_3.}
	
	\bibitem{ArtemPPN:2017}
	D.A. Artemenkov, A. A. Zaitsev, P. I. Zarubin, Phys. Part. Nucl. \textbf{48} 147(2017); \href{https://doi.org/10.1134/S1063779617010026}{DOI: 10.1134/S1063779617010026.}
	
	\bibitem{ArtemPAN:2017}
	D.A. Artemenkov \textit{et al}., Phys. At. Nucl. \textbf{80}, 1126 (2017); \href{https://doi.org/10.1134/S1063778817060047}{DOI:10.1134/S1063778817060047.}
	
	\bibitem{BelagaPAN:1995}
	V.V. Belaga, A.A. Benjaza, V.V. Rusakova, D.A. Salomov, G.M. Chernov, Phys. At. Nucl. \textbf{58}, 1905 (1995); \href{https://arxiv.org/abs/1109.0817}{arXiv:1109.0817.}
	
	\bibitem{AndreevaPAN:1996}
	N.P. Andreeva \textit{et al}., Phys. At. Nucl. \textbf{59}, 102 (1996); \href{https://arxiv.org/abs/1109.3007}{arXiv:1109.3007.}

	\bibitem{ArtemRM:2018}
	D.A. Artemenkov \textit{et al}., Rad. Meas. \textbf{119}, 199 (2018); \href{https://doi.org/10.1016/j.radmeas.2018.11.005}{DOI: 10.1016/j.radmeas.2018.11.005.}
	
	\bibitem{ArtemFewBP:2020}
	D.A. Artemenkov \textit{et al}., Springer Proc. Phys. \textbf{238}, 137 (2020);\\ \href{https://doi.org/10.1007/978-3-030-32357-8_24}{DOI: 10.1007/978-3-030-32357-8\_24.}
	
	\bibitem{ArtemEPJ:2020}
	D.A. Artemenkov \textit{et al}., Eur. Phys. J. A \textbf{56}, 250 (2020);\\ \href{https://doi.org/10.1140/epja/s10050-020-00252-3}{DOI: 10.1140/epja/s10050-020-00252-3.}
	
	\bibitem{AndreevaSovJNP:1988}
	N.P. Andreeva \textit{et al.}, Sov. J. Nucl. Phys. \textbf{47} 102 (1988); Yad. Fiz. \textbf{47} 157 (1988) and Dubna JINR - 86-828.
	
	\bibitem{NaghyJPG:1988}
	A. El-Naghy \textit{et al.}, J. Phys. G, \textbf{14} 1125 (1988); \href{https://doi.org/10.1088/0305-4616/14/8/015}{DOI:10.1088/0305-4616/14/8/015 }
	
	\bibitem{AdamovichPRC:1989}
	M.I. Adamovich \textit{et al.}, Phys. Rev. C \textbf{40}, 66 (1989); \href{https://doi.org/10.1103/PhysRevC.40.66}{DOI: 10.1103/PhysRevC.40.66}
	
	\bibitem{AdamovichZPA:1995}
	M.I. Adamovich \textit{et al.}, Z. Phys. A \textbf{351}, 311 (1995); \href{https://doi.org/10.1007/BF01290914}{DOI: 10.1007/BF01290914.}
	
	\bibitem{AdamovichEPJA:1999}
	M.I. Adamovich \textit{et al.}, Eur. Phys. J. A \textbf{5}, 429 (1999); \href{https://doi.org/10.1007/s100500050306}{DOI: 10.1007/s100500050306.}
	
	\bibitem{BecquerelWeb}
	The BECQUEREL Project \href{http://becquerel.jinr.ru/movies/movies.html}{http://becquerel.jinr.ru/movies/movies.html.}
	
	\bibitem{GlagolevEPJA:2001}
	V.V. Glagolev \textit{et al.}, Eur. Phys. J. A \textbf{11}, 285 (2001); \href{https://doi.org/10.1007/s100500170067}{DOI: 10.1007/s100500170067.}
	
	\bibitem{KrasnovCzech:1996}
	S.A. Krasnov \textit{et al.}, Czechoslovak J. of Phys. \textbf{46} 531 (1996); \href{https://doi.org/10.1007/BF01690674}{DOI: 10.1007/BF01690674}

	
\end{thebibliography}

\end{document}